\documentclass[aps,prc,twocolumn,fleqn]{revtex4}

\usepackage{graphicx}
\usepackage{amsmath,amssymb,amsopn}
\usepackage{color}

\newcommand{\be}{\begin{equation}}
\newcommand{\ee}{\end{equation}}
\newcommand{\ba}{\begin{eqnarray}}
\newcommand{\ea}{\end{eqnarray}}
\newcommand{\ban}{\begin{eqnarray*}}
\newcommand{\ean}{\end{eqnarray*}}
\newcommand \nn {\nonumber}
\newcommand{\sla}{\!\!\!/ \,}
\vspace{4cm}
\begin{document}

\title{Momentum Broadening of a Fast Parton in a Perturbative Quark-Gluon Plasma}

\author{Abhijit Majumder}

\affiliation{Department of Physics, The Ohio State University, Columbus, OH 43210,  USA}

\author{Berndt M\"uller}

\affiliation{Department of Physics, Duke University, Durham, NC 27708, USA}

\author{Stanis\l aw Mr\' owczy\' nski}

\affiliation{Institute of Physics, Jan Kochanowski University, 
25-406 Kielce, Poland}
\affiliation{So\l tan Institute for Nuclear Studies, 00-681 Warsaw, Poland}

\date{November 17, 2009}

\begin{abstract}

The average transverse momentum transfer per unit path length to a fast 
parton scattering elastically in a perturbative quark-gluon plasma is 
related to the radiative energy loss of the parton. We first calculate 
the momentum transfer coefficient $\hat q$ in terms of a classical 
Langevin problem and then define it quantum-mechanically through 
scattering matrix element. After treating the well known case of a 
quark-gluon plasma in equilibrium we consider an off-equilibrium 
unstable plasma. As a specific example, we treat the two-stream 
plasma with unstable modes of longitudinal chromoelectric field. In 
the presence of the instabilities, $\hat q$ is shown to exponentially 
grow in time.

\end{abstract}

\maketitle

\section{Introduction}

When a highly energetic parton travels through dense QCD matter, 
it receives random kicks from elastic interactions with  
constituents of the plasma. The average transverse momentum 
transfer per unit path length is related to the radiative energy 
loss of the parton \cite{Baier:1996sk}. The parameter describing 
the average amount of transverse momentum broadening per unit 
length is called $\hat{q}$ and is defined as
\be
\hat{q} \equiv d\langle\Delta {\bf p}_T^2\rangle / dz ,
\label{qhat}
\ee
when the fast parton flies along the direction $z$. The values 
of $\hat{q}$ extracted from experimental data on relativistic 
heavy-ion collisions vary in a rather broad range 
$0.5-15\;{\rm GeV^2/fm}$ depending on the model of hard particle 
propagation in strongly interacting matter produced in nuclear 
collisions \cite{Eskola:2004cr,Baier:2006fr}. 

The calculations of $\hat{q}$ for the case of perturbative 
quark-gluon plasma in equilibrium are well understood (see 
\cite{Arnold:2008vd,Peigne:2008wu} for recent work).  For such 
a plasma the value of $\hat{q}$ is predicted to lie at the lower
end of the range of values deduced from experiments
\cite{Baier:2006fr}. However, it is not at all obvious that the 
quark-gluon plasma produced in a heavy ion collision is in 
equilibrium during the whole time interval of the propagation 
of the fast parton. It is rather expected that the momentum 
distribution of plasma partons is anisotropic at the early 
stage of relativistic heavy-ion. And recently $\hat{q}$ has been 
computed \cite{Romatschke:2006bb,Baier:2008js} for the case of 
homogenous quark-gluon plasma with anisotropic momentum 
distribution. However, such a plasma is unstable due to the 
chromomagnetic modes (for a review see \cite{Mrowczynski:2005ki}). 
The fact that unstable systems are explicitly time dependent as 
unstable modes exponentially grow in time has not been taken into 
account in the previous analytical calculations 
\cite{Romatschke:2006bb,Baier:2008js}. However, numerical 
simulations \cite{Dumitru:2007rp,Schenke:2008gg} indicate that 
$\hat{q}$ receives a sizable contribution from these unstable 
modes. Also, possible phenomenological consequences of the momentum
broadening effect of an unstable plasma on hard partons have been 
proposed \cite{Dumitru:2007rp,Majumder:2006wi}.

In this paper we develop the approach suggested in \cite{Majumder:2007hx} 
(see also \cite{Liang:2008vz}) to study the transverse momentum 
fluctuations in terms of classical Langevin problem. 
We calculate $\hat{q}$ by treating the parton as an energetic 
classical particle with SU(3) color charge moving in the presence 
of the fluctuating color fields. Then, $\hat{q}$ is expressed 
through the correlation function of chromodynamic fields computed 
in \cite{Mrowczynski:2008ae}, or equivalently, through by the 
unordered gluon propagator in Hard Thermal Loop (HTL) 
approximation. For the equilibrium plasma the Langevin approach 
recovers the known result obtained within the standard thermal 
field theory \cite{lebellac}. We then apply the Langevin approach  
to the case of a parton traversing an off-equilibrium unstable plasma. 
For the sake of simplicity, we study the two-stream system and 
consider only the contribution from the unstable modes of longitudinal 
chromoelectric field, neglecting the contribution from transverse 
fields. We show that $\hat q$ grows exponentially in time due to 
the exponential growth of the unstable mode amplitude.

\section{Classical approach}
\label{sec-classical}

We consider a classical parton which moves across a quark-gluon 
plasma.  Its motion is described by the Wong equations 
\cite{Wong:1970fu}
\be
\label{EOM-1a}
\frac{d x^\mu(\tau)}{d \tau} = u^\mu(\tau ) ,
\ee
\be
\label{EOM-1b}
\frac{d p^\mu(\tau)}{d \tau} = g Q^a(\tau ) \, F_a^{\mu \nu}\big(x(\tau )\big) 
\, u_\nu(\tau ) ,
\ee
\be
\label{EOM-1c}
\frac{d Q_a(\tau)}{d \tau} = - g f^{abc} p_\mu (\tau ) \, 
A^\mu _b \big(x(\tau )\big) \, 
Q_c(\tau) ,
\ee
where $\tau$, $x^\mu(\tau )$, $u^\mu(\tau)$ and  $p^\mu(\tau)$
are, respectively, the parton's  proper time, its trajectory, 
four-velocity and  four-momentum; $F_a^{\mu \nu}$ and  
$A_a^\mu$ denote the chromodynamic field strength tensor and
four-potential, respectively, and $Q^a$ is the classical 
color charge of the parton.

We look for a solution of the Wong equations in a specific
gauge and the gauge dependence of our results is discussed 
{\it a posteriori}. We assume that the potential vanishes along 
the parton's trajectory by imposing the axial gauge condition
$p_\mu (\tau ) \, A^\mu _a \big(x(\tau )\big) = 0$.
Then, Eq.~(\ref{EOM-1c}) simply states that $Q_a$ is constant
as a function of $\tau$. 

We solve Eqs.~(\ref{EOM-1a}, \ref{EOM-1b}) assuming that the 
momentum of the parton is so high that the changes in its direction 
of motion caused by the interactions with the medium are small 
compared with the momentum of the parton: 
$|\Delta {\bf p}|/|{\bf p}| \ll 1$. The  change in the velocity 
vector ${\bf v}$ of the parton are then negligible, and we can 
consider the parton to move along a straight-line path with 
constant velocity. If the parton is massless or nearly massless 
(a light quark or a gluon), its speed can be approximated by the 
speed of light. Although we assume further on that the fast parton 
is at least approximately massless, we note that the approximation 
of a straight line trajectory with constant velocity also 
applies to a heavy quark with mass $M$ and momentum such 
that $|{\bf p}| = \gamma M|{\bf v}| \gg |\Delta {\bf p}|$.

In the limit of massless particle, the proper time becomes 
a meaningless concept and we formally solve the equation of 
motion (\ref{EOM-1b}) to eliminate it. Then, we get
\be
\label{EOM-3}
p^\mu(\tau ) = p^\mu(0) + g Q_a \int_0^\tau d\tau' 
F_a^{\mu \nu}\big(x(\tau' )\big) u_\nu(\tau' ) .
\ee
In general, this is not an explicit solution of (\ref{EOM-1b}) 
as the four-velocity $u_\nu(\tau)$ on the right-hand side of 
(\ref{EOM-3}) depends on the four-momentum. However, the 
four-velocity, as explained above, is approximated by 
a constant: $u_\nu(\tau) \approx u_\nu(0)$. 
Assuming that the direction of motion of the parton is in the 
{\em positive} $x^3$-direction, we have explicitly 
$u^\mu = \gamma (1,0,0,v)$ with the Lorentz factor 
$\gamma = (1 - v^2)^{-1/2}$. Then (\ref{EOM-3}) becomes 
\be
\label{EOM-4}
p^\mu(t) = p^\mu(0) + g Q_a \int_0^t dt' 
\left(F_a^{\mu 0}(x') - v F_a^{\mu 3}(x') \right) ,
\ee
where $x' \equiv x(t')$ and $t=\gamma\tau$.

In the limit $v\to 1$, the form of (\ref{EOM-4}) motivates the 
introduction of light-cone variables defined as
\be
p^{\pm} \equiv \frac{p^0 \pm p^3}{\sqrt{2}} .
\ee
In the new coordinates the four-vector components are understood 
as $p^{\mu} =(p^+,p^-,p^1,p^2)$ and the product of two 
four-vectors is
\be
p^\mu q_\mu = p^+ q^- + p^-q^+ - p^1 q^1 - p^2 q^2 .
\ee
Assuming that the parton's initial position is ${\bf x}(0)=0$, 
its trajectory is given by $x^\mu(t) = \sqrt{2}(t,0,0,0)$, and 
Eq.~(\ref{EOM-4}) takes the form
\be
\label{EOM-5}
p^\mu(x^+) = p^\mu(0) + g Q_a \int_0^{x^+} dy^+ 
F_a^{\mu -}(y^+) .
\ee
Here $F_a^{\mu -}(y^+)$ is a short-hand notation for 
$F_a^{\mu -}\big(x^\nu(y^+)\big)$ with $x^\nu(y^+)=(y^+,0,0,0)$.
If the parton moves in the {\em negative} $x^3$-direction, 
Eq.~(\ref{EOM-5}) changes to 
\be
\label{EOM-6}
p^\mu(x^-) = p^\mu(0) + g Q_a 
\int_0^{x^-} dy^- F_a^{\mu +}(y^-) .
\ee
In the following we shall assume that the parton travels in 
the {\em positive} $x^3$-direction.

In the spirit of the Langevin approach, we consider the ensemble 
average $\langle p^\mu(x^+) p^\nu(x^+) \rangle$ indicated by
the angular brackets. The ensemble average involves averaging
over color charges. We recall that 
\be
\int dQ \,Q_a Q_b = C_2 \delta^{ab},
\ee
where $C_2 = 1/2$ for particles (quarks) in fundamental 
representation of the ${\rm SU}(N_c)$ group and $C_2 = N_c$ 
for particles (gluons) in adjoint representation. Then, we find
\begin{multline}
\label{p^mu-p^nu-1}
\langle p^\mu(x^+) p^\nu(x^+) \rangle 
= p^\mu(0) p^\nu(0) 
\\ 
+ g^2 C \int_0^{x^+} dy^+_1 \, dy^+_2 \;
\langle F_a^{\mu -}(y^+_1) F_a^{\nu -}(y^+_2) \rangle \,,
\end{multline}
where the color factor $C$ is given as
\be
C \equiv \left\{
\begin{array}{ccl}
\frac{C_2}{N_c}= \frac{1}{2N_c}
& & {\rm (fundamental~rep.)}  
\\[3mm]
\frac{C_2}{N_c^2-1} = \frac{N_c}{N_c^2-1}
&  & {\rm (adjoint~rep.)} .
\end{array}
\right.
\ee
It is related to the eigenvalue $C_R$
($R=F,A$) of the quadratic Casimir operator as
\be
\label{casimir}
C_R = C (N_c^2 -1) .
\ee

Let us consider $\langle p^\mu(x^+) p_\mu(x^+) \rangle$
in more detail:
\begin{multline}
\label{p^mu-p_mu-1}
\langle p^\mu(x^+) p_\mu(x^+) \rangle = 
p^\mu(0) p_\mu(0) 
\\
+ g^2 C \int_0^{x^+} dy^+_1\, dy^+_2 \; g_{\mu\nu}
\langle F_a^{\mu -}(y^+_1) F_{a}^{\nu -}(y^+_2)\rangle .
\end{multline}
An interesting feature of (\ref{p^mu-p_mu-1}) is that although
$\mu$ runs over all four components $0,1,2,3$ or $+,-,1,2$, 
only the transverse components ($\mu = 1,2$) contribute to 
$g_{\mu\nu}\langle F_a^{\mu -}(y^+_1) F_{a}^{\nu -}(y^+_2) \rangle$. 
This is easily seen by writing the expression out in light-cone 
coordinates and recalling that the antisymmetry of $F_a^{\alpha\beta}$ 
requires $F_a^{- -}=0$. The terms for $\mu=+,-$ thus vanish and
consequently,
\begin{multline}
g_{\mu\nu}F_a^{\mu -}(y_1) F_{a}^{\nu -}(y_2)
= \\
- F_a^{1 -}(y_1) F_a^{1 -}(y_2) 
- F_a^{2 -}(y_1) F_a^{2 -}(y_2) .
\end{multline}
The fact that only the transverse components ($\mu = 1,2$) 
contribute implies that we can write
\begin{multline}
\label{dp_T-dp_T-1}
\langle \Delta {\bf p}_T^2(x^+)\rangle \equiv
\langle \Delta {\bf p}_T(x^+) \cdot 
\Delta {\bf p}_T(x^+) \rangle
\\
= - g^2 C \int_0^{x^+} dy^+_1\, dy^+_2 \; g_{\mu\nu}
\langle F_a^{\mu -}(y^+_1) F_{a}^{\nu -}(y^+_2) \rangle ,
\end{multline}
where $\Delta{\bf p}_T(x^+) = {\bf p}_T(x^+) - {\bf p}_T(0)$. 

Equation (\ref{dp_T-dp_T-1}), which is the main result of this section,
has been derived in a specific gauge and the right-hand side 
of the equation is, in general, gauge dependent. The gauge
independence can be restored by inserting the Wilson line
between the fields  $F_a^{\mu -}(y^+_1)$ and 
$F_{a}^{\nu -}(y^+_2)$ in the correlation function 
$\langle F_a^{\mu -}(y^+_1) F_{a}^{\nu -}(y^+_2) \rangle$.
However, the correlation functions, which are further used
in our study, are computed in the Hard Loop approximation.
These functions appear to be gauge independent even without
the Wilson-line insertion, as discussed in detail in Sec.~VIIIA
of \cite{Mrowczynski:2008ae}. Therefore, Eq.~(\ref{dp_T-dp_T-1})
is gauge independent within the approximations we apply. 

We now assume that the quark-gluon plasma is translationally 
invariant in space and time and, hence, the correlator 
$\langle F_a^{\mu -}(y^+_1) F_{a}^{\nu -}(y^+_2) \rangle$ 
depends only on the difference $y^+_1 - y^+_2$. This assumption,
which is relevant for equilibrium plasmas, will be not adopted
for the two-stream system discussed in Sec.~\ref{sec-2-streams} 
as the growth of unstable modes break the translational 
invariance in time. Making use of the translational invariance, 
we introduce the fluctuation spectrum
\be
\langle F_a^{\mu -} F_{a}^{\nu -} \rangle_k
\equiv \int d^4x e^{ikx} 
\langle F_a^{\mu -}(x) F_{a}^{\nu -}(0) \rangle \,,
\ee
which allows us to express Eq.~(\ref{dp_T-dp_T-1}) as 
\begin{multline}
\label{dp_T^2-1}
\langle \Delta {\bf p}_T^2(x^+)\rangle
= - g^2 C \int_0^{x^+} dy^+_1\, dy^+_2 
\\
\times
\int \frac{d^4k}{(2\pi)^4} \; e^{ik^-(y^+_1 - y^+_2)}
g_{\mu \nu}\langle F_a^{-\mu} F_a^{\nu -} \rangle_k .
\end{multline}
Since the double integral over $y^+_1$ and $y^+_2$ tends 
to a delta function of $k^-$ in the long-time limit
\begin{multline}
\int_0^{x^+} dy^+_1\, dy^+_2 \; e^{ik^-(y^+_1 - y^+_2)} 
\\
= \frac{4 \sin^2(k^- x^+/2)}{(k^-)^2} 
\buildrel{x^+ \rightarrow \infty}\over{\longrightarrow} 
2 \pi x^+ \delta (k^-) ,
\end{multline}
we can define the transport coefficient
\begin{multline}
\label{qhat-formula}
\hat q  \equiv  \lim_{t \rightarrow \infty}
\frac{1}{t} \langle \Delta {\bf p}_T^2(t)\rangle 
= \lim_{x^+ \rightarrow \infty}
\frac{\sqrt{2}}{x^+} \langle \Delta {\bf p}_T^2(x^+)\rangle 
\\
= - 2 \sqrt{2} \pi g^2 C 
\int \frac{d^4k}{(2\pi)^4} \; \delta (k^-) \: g_{\mu\nu}
\langle F_a^{\mu -} F_{a}^{\nu -} \rangle_k ,
\end{multline}
which was  introduced in Eq.~(\ref{qhat}) as the transverse 
momentum broadening per unit path length.

From here on we can proceed in two different ways. We can 
either use the fluctuation spectra of chromodynamic fields 
derived in \cite{Mrowczynski:2008ae}, or we can express 
$\langle F_a^{\mu -} F_{a}^{\nu -} \rangle_k$ through
the unordered gluon propagator in the HTL  approximation. 
In the two following sections we pursue both approaches. 
The first one is shorter, while the second one allows us to 
relate our results to those obtained in \cite{Baier:2008js}.

\section{Fluctuation spectra of chromodynamic fields}
\label{sec-field-fluctuations}

The fluctuation spectra of the equilibrium plasma, as computed in
\cite{Mrowczynski:2008ae}, are
\ba
\langle E^i_a E^j_b\rangle_k
&=& 
2 \delta^{ab} T \omega^3 \bigg[
\frac{k^ik^j}{{\bf k}^2}
\frac{\Im \varepsilon_L(\omega,{\bf k})}
{|\omega^2 \varepsilon_L(\omega,{\bf k})|^2}
\nn \\
&+&
\Big(\delta^{ij} - \frac{k^ik^j}{{\bf k}^2}\Big)
\frac{\Im \varepsilon_T(\omega,{\bf k})}
{|\omega^2 \varepsilon_T(\omega,{\bf k})-{\bf k}^2|^2}
\bigg] ,
\nn \\[2mm]
\label{BiBj-spec-eq}
\langle B^i_a B^j_b\rangle_k
&=& 2 \delta^{ab} T \omega \,{\bf k}^2
 \Big(\delta^{ij} - \frac{k^i k^j}{{\bf k}^2} \Big) 
\nn \\
&\times& \frac{\Im \varepsilon_T (\omega,{\bf k})}
{|\omega^2 \varepsilon_T(\omega,{\bf k})-{\bf k}^2|^2} ,
\nn \\[2mm]
\label{BiEj-spec-eq}
\langle B^i_a E^j_b\rangle_k
&=& \langle E^j_a B^i_b\rangle_k
=2 \delta^{ab} T \omega^2 \epsilon^{ikj}k^k
\nn \\
&\times&
\frac{\Im \varepsilon_T (\omega,{\bf k})}
{|\omega^2 \varepsilon_T(\omega,{\bf k})-{\bf k}^2|^2} ,
\ea
where $k=(\omega, {\bf k})$, $\epsilon^{ijk}$ is the antisymmetric
tensor, $T$ is the system's temperature and 
$\varepsilon_{L,T}(\omega,{\bf k})$ are choromodielectric 
functions which for the equilibrium plasma of massless particles 
are well-known to be \cite{lebellac}
\ba
\label{Re-Im-eL-massless}
\Re\varepsilon_L(\omega,{\bf k}) 
&=& 
1+ \frac{m_D^2}{{\bf k}^2}
\bigg[
1 - \frac{\omega}{2|{\bf k}|}
{\rm ln}\bigg|\frac{\omega + |{\bf k}|}{\omega - |{\bf k}|} \bigg| 
\bigg] , 
\nn \\[2mm]
\Im\varepsilon_L(\omega,{\bf k}) 
&=& \frac{\pi}{2} \: 
\Theta ({\bf k}^2 -\omega^2) \:
\frac{m_D^2 \omega}{|{\bf k}|^3} ,
\ea
\ba
\label{Re-Im-eT-massless}
\Re\varepsilon_T(\omega,{\bf k}) 
&=& 
1 - \frac{m_D^2}{2{\bf k}^2}\bigg[
1 - \frac{\omega^2 - {\bf k}^2}{2\omega |{\bf k}|}
{\rm ln}\bigg|\frac{\omega + |{\bf k}|}{\omega - |{\bf k}|} \bigg|
\bigg] , 
\nn \\[2mm]
\Im\varepsilon_T(\omega,{\bf k}) 
&=& \frac{\pi}{4} \:
\Theta ({\bf k}^2 -\omega^2) \:
\frac{m_D^2({\bf k}^2- \omega^2)}{\omega |{\bf k}|^3} ,
\ea
where $m_D$ is the Debye mass:
\be
\label{m_D}
m_D^2 = \frac{g^2 T^2}{6}\, (N_f + 2N_c) .
\ee

Since
\ba
- g_{\mu\nu} \langle F_a^{\mu -} F_{a}^{\nu -} \rangle_k
&=& \langle E^x_a E^x_a \rangle_k 
 +  \langle E^y_a E^y_a \rangle_k 
\nn \\
&-& \langle E^x_a B^y_a \rangle_k
 + \langle E^y_a B^x_a \rangle_k
\nn \\
&-&  \langle B^y_a E^x_a \rangle_k
 +  \langle B^x_a E^y_a \rangle_k
\nn \\ 
&+& \langle B^x_a B^x_a \rangle_k
 +  \langle B^y_a B^y_a \rangle_k ,
\ea
Equation (\ref{qhat-formula}) yields
\ba
\label{qhat-class}
\hat q &=&
2 g^2 C_R T \int \frac{d^3k}{(2\pi)^3} \: 
\frac{k_T^2}{k_z {\bf k}^2} 
\nn \\
&& \times \left[ \frac{\Im \varepsilon_L(k_z,{\bf k})}
{|\varepsilon_L(k_z,{\bf k})|^2} 
+ \frac{ k_z^2 k_T^2 \: \Im \varepsilon_T(k_z,{\bf k})}
{| k_z^2 \varepsilon_T(k_z,{\bf k}) - {\bf k}^2|^2} \right] .
\ea
The classical formula (\ref{qhat-class}) holds for the inverse 
wave vectors of the fields which are much longer than the de 
Broglie wavelength of plasma particles. It requires 
$|{\bf k}| \ll T$ where $T$ is the plasma temperature. 
For larger wave vectors a quantum approach is needed which 
is discussed in Sec.~\ref{sec-quantum-qhat}. 

Keeping in mind that the equations 
$$
\varepsilon_L(\omega,{\bf k}) = 0 ,\;\;\;\;\;
\omega^2 \varepsilon_T(\omega,{\bf k}) - {\bf k}^2 = 0,
$$
determine the longitudinal and transverse collective modes in 
the isotropic plasma, one sees that the transverse momentum
broadening, as given by the formula (\ref{qhat-class}), is 
determined by the interaction of the fast parton with plasma 
collective excitations.

\section{Gluon propagator}
\label{sec-propagator}

As already mentioned, 
$\langle F_a^{\mu -} F_{a}^{\nu -} \rangle_k$, which 
enters Eq.~(\ref{qhat-formula}), can be expressed through 
the unordered gluon propagator in the HTL approximation.

Neglecting all non-Abelian contributions, the Fourier transform 
of the field strength tensor is expressed through the 
four-potential as 
\be
F_a^{\mu \nu}(k) = -ik^\mu A_a^\nu(k) + i k^\nu A_a^\mu(k) ,
\ee
allowing us to rewrite (\ref{qhat-formula}) as 
\ba
\label{qhat1}
\hat q = 
\sqrt{2} \, \pi g^2 C
\int \frac{d^4k}{(2\pi)^4} \; \delta (k^-) \:
k_T^2 \, \langle A_a^- A_a^- \rangle_k .
\ea
Here we made use of the relation 
$\partial^0 \pm \partial^3 = \sqrt{2} \partial^\mp$
and further used the fact that only the $\mu=1,2$ components 
contribute in the field strength correlator. We also 
dropped all terms proportional to $k^-$ in view of the 
delta function. 

Because $y^+_1$ and $y^+_2$ in Eq.~(\ref{dp_T-dp_T-1}) run 
independently from 0 to $x^+$, the correlator  
$\langle A_a^- A_a^- \rangle_k$ is the Fourier transform 
of the time-unordered gluon Green function. 
Since the plasma of interest is on average color neutral,
the gluon propagators such as $\langle A^\mu_a(x) A^\nu_b(y)\rangle$
are proportional to $\delta^{ab}$. We drop for now 
the color indices but when the gluon propagator 
$\langle A_a^- A_a^- \rangle_k$ is substituted in Eq.~(\ref{qhat1}), 
the coefficient $\delta^{aa} = N_c^2 -1$ will included. 

For translationally invariant systems, one defines the 
time-unordered ($>,<$) Green functions
\ba
iD^>_{\mu \nu}(x-y) &\equiv& \langle A_\mu(x) A_\nu(y) \rangle ,
\\[2mm]
iD^<_{\mu \nu}(x-y) &\equiv& \langle A_\nu(y) A_\mu(x)\rangle ,
\ea
where, as already mentioned, the color indices are suppressed. 
We also introduce the spectral function
\ba
\label{spectral}
\rho_{\mu \nu}(x-y) 
&=& \left\langle [A_\mu(x), A_\nu(y)] \right\rangle 
\nn \\
\label{spec->-<}
&=& iD^>_{\mu \nu}(x-y) - iD^<_{\mu \nu}(x-y) .
\ea

The equilibrium Green functions $D^>_{\mu \nu}(x)$,
$D^<_{\mu \nu}(x)$ obey the Kubo-Martin-Schwinger condition
\be
D^>_{\mu \nu}(t,{\bf x}) = D^<_{\mu \nu}(t+i\beta,{\bf x}),
\ee
or
\be
\label{KMS}
D^>_{\mu \nu}(k) = e^{\beta \omega} \; D^<_{\mu \nu}(k) ,
\ee
where $\beta \equiv 1/T$. Combining the definition 
(\ref{spec->-<}) and the KMS condition (\ref{KMS}), we have 
\ba
\label{<-spec}
iD^<_{\mu \nu}(k) &=& n(\omega) \; \rho_{\mu \nu}(k) ,
\\[2mm]
\label{>-spec}
iD^>_{\mu \nu}(k) &=& \big(n(\omega) + 1\big) \; \rho_{\mu \nu}(k) ,
\ea
where $ n(\omega)\equiv (e^{\beta \omega}-1)^{-1}$ denotes the 
Bose distribution. 

The spectral function in the HTL approximation
can be expressed as \cite{lebellac}
\ba
\label{spec-HTL}
\rho_{\mu \nu}(k) 
&=& \frac{2\, \Im \varepsilon_L(\omega,{\bf k})}
{k^2 |\varepsilon_L(\omega,{\bf k})|^2} 
\: P^L_{\mu \nu}(k) 
\nn \\
&& + \frac{2\, \omega^2 \Im \varepsilon_T(\omega,{\bf k})}
{|\omega^2 \varepsilon_T(\omega,{\bf k}) - {\bf k}^2|^2}
\: P^T_{\mu \nu}(k) ,
\ea
with the projectors defined as 
\ba
\label{P_T}
P_T^{ij}(k) &=& \delta^{ij} - \frac{k^i k^j}{{\bf k}^2} , 
\;\;\;\;\;\;
P_T^{0 \mu}(k) = 0 ,
\\
\label{P_L}
P_L^{\mu \nu}(k) &=& -g^{\mu \nu} +  \frac{k^\mu k^\nu}{ k^2} 
- P_T^{\mu \nu }(k) .
\ea

We identify the correlator $\langle A_a^- A_b^-\rangle_k$ 
from Eq.~(\ref{qhat-formula}) with the time-unordered Green 
function $D_>^{- -}(k)$. Substituting $D_>^{- -}(k)$ given
by Eqs.~(\ref{<-spec}, \ref{spec-HTL}) into Eq.~(\ref{qhat1}), 
we obtain
\ba
\hat q 
&=& 
4 \sqrt{2} \,\pi g^2 C_R 
\int \frac{d^4k}{(2\pi)^4} \; \delta (k^-) \: 
n(\omega) \: k_T^2
\\ \nn
&& \times \left[ \frac{\Im \varepsilon_L(\omega,{\bf k})}
{k^2 |\varepsilon_L(\omega,{\bf k})|^2} 
\: P_L^{--}(k) \right. 
\\ \nn
&& 
+ \left. \frac{\omega^2 \Im \varepsilon_T(\omega,{\bf k})}
{|\omega^2 \varepsilon_T(\omega,{\bf k}) - {\bf k}^2|^2}
\: P_T^{--}(k)
\right] ,
\ea
where $C_R$ is the eigenvalue of the quadratic Casimir operator 
given by Eq.~(\ref{casimir}). 

With the help of formulas (\ref{P_T}, \ref{P_L}) we find
\ba
P_T^{--}(k) 
&=&  \frac{k_T^2}{2{\bf k}^2} ,
\\
P_L^{--}(k) &=& \frac{k^- k^-}{{\bf k}^2} - \frac{k_T^2}{2{\bf k}^2} ,
\ea
and thus, we finally obtain
\ba
\label{qhat-final}
\hat q 
&=& 
2 g^2 C_R \int \frac{d^3k}{(2\pi)^3} \: n(k_z) \: 
\frac{k_T^2}{{\bf k}^2}
\nn \\
&& \times \left[ \frac{\Im \varepsilon_L(k_z,{\bf k})}
{|\varepsilon_L(k_z,{\bf k})|^2} 
+ \frac{ k_z^2 k_T^2 \: \Im \varepsilon_T(k_z,{\bf k})}
{| k_z^2 \varepsilon_T(k_z,{\bf k}) - {\bf k}^2|^2} \right] ,
\ea
where $k_z \equiv k^3$. In the classical field limit, where 
$n(\omega) \approx T/\omega$, we obtain again the expression 
(\ref{qhat-class}).

Since the HTL approximation holds for $|{\bf k}| \ll T$,
the integration in Eq.~(\ref{qhat-final}), as well as in 
the fully classical formula (\ref{qhat-class}), should
be cut off at $|{\bf k}_{\rm max}| \ll T$. For larger wave 
vectors we need a quantum approach, which is discussed in 
the next section.

\section{Quantum approach}
\label{sec-quantum-qhat}

Here we sketch the derivation of $\hat q$  following 
the strategy outlined in \cite{lebellac} for 
the electromagnetic energy loss of a fast muon. We present
how to obtain two contributions to $\hat q$ coming from soft 
($|{\bf q}| \ll T$) and hard ($|{\bf q}| \sim T$) momentum 
transfers. We discuss only the main steps of the derivation
as the procedure is mostly described in the literature
\cite{Arnold:2008vd,Baier:2008js}. This section is 
included for completeness of our study.

Within a quantum approach $\hat q$ is defined as
\be
\label{qhat-quant}
\hat q \equiv \int d^2q_T q_T^2 \frac{d^2 \Gamma}{d^2q_T} \;
\ee
where $\Gamma$ is the rate (probability per unit time) for 
elastic collisions of parton with plasma particles. For a fast parton 
with four-momentum $p_1$ which scatters on plasma constituents of 
species $i$ with four-momenta $p_2$, the rate corresponding 
to the binary process $p_1 + p_2 \to p_1'+ p_2'$  is given as
\begin{eqnarray}
\label{Gamma}
\Gamma &=& \frac{1}{E_1}
\int \frac{d^3p_1'}{2E_1'(2\pi)^3}
\frac{d^3p_2'}{2E_2'(2\pi)^3} 
\frac{d^3p_2}{E_2(2\pi)^3}\,
\\ & &
\times (2\pi )^4\delta^{(4)}( p_1 + p_2 - p_1' - p_2')
\nonumber \\ 
& & \times \sum_i
f_i({\bf p}_2) 
\big[1 \pm f_i({\bf p}_2')\big]
\big| {\cal M}_i( p_1, p_2 ; p_1',p_2') \big|^2 ,
\nonumber
\end{eqnarray}
where $f_i({\bf p})$ is the distribution function of particles
of species $i$ and ${\cal M}_i( p_1, p_2 ; p_1',p_2')$ is 
the scattering matrix element. The definition of $\big|{\cal M}_i\big|^2$ 
involves summation over final spins and colors and an average over 
the initial spin and color of the fast parton. 

The matrix element of parton-parton scattering is assumed to be
dominated by the one-gluon exchange process. Then, the matrix 
element squared can be expressed as the self energy of the fast 
quark. When the fast parton is a massless quark, one finds 
\cite{Baier:2008js}
\be
\label{qhat-sigma-w}
\hat q =  \frac{1}{4 E}
{\rm Tr} \big[ p\sla \Sigma^>_{\rm w}(p)\big] \,
\ee
where $\Sigma^>_{\rm w}(p)$ is the fast quark self-energy 
`weighted' with $q_T^2$. It is given as 
\be
\label{sigma-1-loop}
\Sigma^>_{\rm w}(p)
=  g^2 C_F \int \frac{d^3p'}{2E'(2\pi)^3} \, q_T^2
\gamma^\mu p\sla\!' \gamma^\nu D^<_{\mu \nu}(q) ,
\ee
where $p' = p + q$ and $D^<_{\mu \nu} (q)$ is the gluon 
propagator (stripped of the color factor $\delta^{ab}$) which 
includes the one-loop correction. Changing the orientation 
of the momentum $q$, the propagator $D^<_{\mu \nu} (q)$ 
(\ref{<-spec}) should be replaced by $D^>_{\mu \nu} (q)$ 
(\ref{>-spec}) but we get the same result as $n(-q_0) +1 
= - n(q_0)$.

In the soft domain, where $|{\bf q}| \ll T$, the one-loop
gluon propagator $D^<_{\mu \nu}(k)$ is obtained in the 
HTL approximation. Substituting $D^<_{\mu \nu}(k)$ provided 
by Eqs.~(\ref{<-spec}, \ref{spec-HTL}) into 
Eq.~(\ref{qhat-sigma-w}), we find the differential rate 
which enters Eq.~(\ref{qhat-quant}) as \cite{Wang:2000uj}
\begin{multline}
\label{Gamma-soft}
\frac{d^2 \Gamma_{\rm soft}}{d^2q_T}
= 
2 g^2 C_R \int^{\Lambda} \frac{dq_z}{(2\pi)^3} \:  
\frac{ n(q_z)}{{\bf q}^2}
 \\[2mm]
\times \left[ \frac{\Im \varepsilon_L(q_z,{\bf q})}
{|\varepsilon_L(q_z,{\bf q})|^2} 
+ \frac{ q_z^2 q_T^2 \: \Im \varepsilon_T(q_z,{\bf q})}
{| q_z^2 \varepsilon_T(q_z,{\bf k}) - {\bf q}^2|^2} \right],
\end{multline}
where the upper cut-off $\Lambda$ obeys $gT < \Lambda < T$.
We have here replaced $C_F$ by $C_R$ as a similar analysis 
can be performed for fast gluons. The expression for $\hat q$ 
given by Eq.~(\ref{qhat-quant}) with (\ref{Gamma-soft}) agrees,
as expected, with Eq.~(\ref{qhat-final}). By explicit evaluation 
one shows \cite{Arnold:2008vd} that the soft collision rate 
(\ref{Gamma-soft}) can be approximated as
\be
\label{Gamma-soft-approx}
\frac{d\Gamma_{\rm soft}}{d^2q_T} 
\approx
g^2 \frac{C_R}{(2\pi)^2} 
\frac{T m_D^2}{q_T^2 (q_T^2+ m_D^2)} .
\ee
The hard ($|{\bf q}| \sim T$) contribution to $\hat q$ can 
be found directly from Eqs.~(\ref{qhat-quant},\ref{Gamma}) 
using the unscreened gluon propagator. The approximate result is 
\cite{Arnold:2008vd} 
\be
\label{Gamma-hard}
\frac{d\Gamma_{\rm hard} }{d^2q_T} 
\approx
g^4 \frac{C_R}{(2\pi)^2} \frac{\rho}{q_T^4}
\ee
where $\rho$ is the properly normalized parton density in the plasma.

The complete result for $\hat{q}$ is obtained by summing up
the hard and soft contributions. An arbitrary parameter $\Lambda$,
which separates the soft momentum transfers from the hard one, is
the ultraviolet cut-off for the soft contribution and the infrared
cut-off for the hard contribution. $\Lambda$ is eliminated from 
the final answer, as the soft contribution depends logarithmically 
on $\Lambda$ and the hard contribution depends logarithmically 
on $\Lambda^{-1}$. Since the hard contribution also depends  
logarithmically on the ultraviolet bound provided by two-body
kinematics of the scattering process, we can simply extrapolate 
the soft contribution to the hard domain with no significant error 
\cite{Arnold:2008vd}. In conclusion, the integral in the soft contribution 
to $\hat{q}$, as given by Eqs.~(\ref{qhat-quant},\ref{Gamma-soft-approx}), 
can be extended to the maximal momentum transfer $q_T^{\rm max}$ 
or $k_T^{\rm max}$ roughly equal $\sqrt{ET}$ where $E=p^0$ is the energy 
of the fast parton.

\section{Two-stream unstable plasma}
\label{sec-2-streams}

Our aim here is to calculate the momentum broadening of a fast parton
in unstable anisotropic plasmas. As mentioned in the Introduction,
this problem was studied previously \cite{Romatschke:2006bb,Baier:2008js}
in a way similar to that presented in Sec.~\ref{sec-quantum-qhat}.
Namely, the hard thermal loop gluon propagator of an equilibrium 
plasma was replaced by the hard loop propagator of an anisotropic 
plasma \cite{Mrowczynski:2004kv}. Such an approach, however, does
not take into account that unstable systems are intrinsically time
dependent because the unstable modes grow exponentially 
in time. We show here how these modes influence the momentum 
broadening of a fast parton. For the sake of analytical tractability, we 
do not consider the anisotropic plasma with a momentum distribution 
likely to be found in the early stage of relativistic heavy-ion collisions, 
as the correlation functions of chromodynamic fields are difficult to
obtain for such a plasma. Instead, we consider a fast parton in the 
more tractable, but also unstable two-stream plasma. With this 
simplified example we hope to elucidate the general features of the 
problem of momentum broadening in unstable plasmas.

The distribution function of the two-stream plasma is chosen
to be of the form
\be
\label{f-2-streams}
f({\bf p}) = (2\pi )^3 n 
\Big[\delta^{(3)}({\bf p} - {\bf p}_0) 
+ \delta^{(3)}({\bf p} + {\bf p}_0) \Big] ,
\ee
where $n$ is the effective parton density in a single stream. The 
distribution function \eqref{f-2-streams} should be treated as an 
idealization of the two-peak distribution where the particles have 
momenta close to ${\bf p}_0$ or $-{\bf p}_0$.

\begin{widetext}

The two-stream plasma is unstable with respect to both longitudinal and 
transverse modes. Because the two-stream plasma does not have a direct
phenomenological application, and we are only interested in showing the
effect of unstable modes on $\hat{q}$ in principle, we consider for simplicity 
that the parton interacts only with the longitudinal electric fields. It is worth
noting that there exist also unstable transverse modes, which will have a
similar effect on $\hat{q}$. The correlation function of the longitudinal field
components was derived in \cite{Mrowczynski:2008ae}. We are mostly 
interested in the exponentially growing modes but the correlation function 
which represents only such modes becomes meaningless when the growth
rate tends to zero. Therefore, we denote the correlation function with the 
subscript label ``exp'' to indicate that it includes both, the exponentially 
growing and the exponentially decaying modes. The expression for the
correlation function reads \cite{Mrowczynski:2008ae}:
\ba
\label{E^iE^j-2-stream}
\langle E_a^i(t_1,{\bf r}_1) E_b^j(t_2,{\bf r}_2) 
\rangle_{\rm exp}
&=& \frac{g^2}{2} \,\delta^{ab} n 
\int \frac{d^3k}{(2\pi)^3} 
\frac{e^{i {\bf k}({\bf r}_1 - {\bf r}_2)}}{{\bf k}^4}
\frac{k^i k^j}{(\omega_+^2 - \omega_-^2)^2}
\frac{
\big(\gamma_{\bf k}^2 + ({\bf k} \cdot {\bf u})^2\big)^2} 
{\gamma_{\bf k}^2}
\nonumber \\[2mm]
&\times& 
\Big[
\big(\gamma_{\bf k}^2 + ({\bf k} \cdot {\bf u})^2\big)
\cosh \big(\gamma_{\bf k} (t_1 + t_2)\big)
+
\big(\gamma_{\bf k}^2 - ({\bf k} \cdot {\bf u})^2\big)
\cosh \big(\gamma_{\bf k} (t_1 - t_2)\big) \Big] ,
\ea
where ${\bf u} \equiv {\bf p}_0/E_{{\bf p}_0}$ is the stream velocity
and $\pm \omega_{\pm}({\bf k})$ are the four roots of the dispersion 
equation $\varepsilon_L(\omega,{\bf k}) = 0$:
\be
\label{roots}
\omega_{\pm}^2({\bf k}) = \frac{1}{{\bf k}^2}
\bigg[{\bf k}^2 ({\bf k} \cdot {\bf u})^2
+ \mu^2 \big({\bf k}^2 - ({\bf k} \cdot {\bf u})^2\big)
\pm \mu \sqrt{\big({\bf k}^2 - ({\bf k} \cdot {\bf u})^2\big)
\Big(4{\bf k}^2 ({\bf k} \cdot {\bf u})^2 +
\mu^2 \big({\bf k}^2 - ({\bf k} \cdot {\bf u})^2\big)\Big)} 
\; \bigg] ,
\ee
\end{widetext}
with $\mu^2 \equiv g^2n/2 E_{{\bf p}_0}$. $\gamma_{\bf k}$ is the 
instability growth rate ($0 < \gamma_{\bf k} \in \mathbb{R}$) defined as 
$\omega_-({\bf k})= i \gamma_{\bf k}$. The integration over 
${\bf k}$ in Eq.~(\ref{E^iE^j-2-stream}) is performed over the 
domain of unstable longitudinal modes defined by the requirement
\be
{\bf k} \cdot {\bf u} \not= 0
\;\;\;
{\rm and}
\;\;\;
{\bf k}^2 ({\bf k} \cdot {\bf u})^2 
< 2 \mu^2 \big({\bf k}^2 - ({\bf k} \cdot {\bf u})^2\big) .
\ee
The correlation function (\ref{E^iE^j-2-stream}) is obviously
invariant with respect to space translations -- it depends on the 
difference $({\bf r}_1 - {\bf r}_2)$ only. The initial plasma state
is on average homogeneous and it remains so over the course of
time. The time dependence of the correlation function 
(\ref{E^iE^j-2-stream}) is very different from the space dependence,
and it is not invariant under time translations. The electric fields 
grow exponentially and so does the correlation function, both in 
the variables $(t_1 + t_2)$ and $(t_1 - t_2)$. The fluctuation spectrum 
also evolves in time as the growth rate of unstable modes is wave 
vector dependent. After a sufficiently long time the fluctuation 
spectrum is dominated by the fastest growing modes.

Substituting the correlation function (\ref{E^iE^j-2-stream}) into
Eq.~(\ref{p^mu-p_mu-1}) and reverting to integration over time $t$
instead of the light-cone variable $y^+$, we find 
\begin{widetext}
\ba
\label{p2-2-stream-1}
\langle \Delta{\bf p}_T^2(t)\rangle 
&=&
g^2 C \int_0^t dt'\int_0^t dt''
\Big[
\langle E^x_a(t',{\bf r}(t')\: E^x_a(t'',{\bf r}(t'')\rangle
+ \langle E^y_a(t',{\bf r}(t')\: E^y_a(t'',{\bf r}(t'')\rangle \Big]
\nn \\[2mm] 
&=&
\frac{g^4}{4} \, C_R \, n 
\int \frac{d^3k}{(2\pi)^3} 
\frac{k_T^2}{{\bf k}^4 (\omega_+^2 - \omega_-^2)^2}
\frac{\big(\gamma_{\bf k}^2 + ({\bf k} \cdot {\bf u})^2\big)^2} 
{\gamma_{\bf k}^2(k_z^2 + \gamma_{\bf k}^2)}
\nn \\[2mm] 
&\times& 
\bigg[
\big(\gamma_{\bf k}^2 + ({\bf k} \cdot {\bf u})^2\big)
\big(|e^{(i k_z + \gamma_{\bf k})t}-1|^2
   + |e^{(i k_z - \gamma_{\bf k})t}-1|^2 \big)
+ 4 \big(\gamma_{\bf k}^2 - ({\bf k} \cdot {\bf u})^2\big)
\frac{k_z^2 - \gamma_{\bf k}^2}{k_z^2 + \gamma_{\bf k}^2}
\bigg] .
\ea
Taking into account only the fastest growing contribution to 
Eq.~(\ref{p2-2-stream-1}), we obtain
\be
\label{qhat-unstable}
\hat{q} = \frac{d\langle \Delta {\bf p}_T^2(t)\rangle}{dt} 
 \approx 
\frac{g^4}{2} \, C_R \, n  
\int \frac{d^3k}{(2\pi)^3} \: e^{2\gamma_{\bf k}t} \:
\frac{k_T^2 \big(\gamma_{\bf k}^2 + ({\bf k} \cdot {\bf u})^2\big)^3}
{{\bf k}^4 (\omega_+^2 - \omega_-^2)^2
\gamma_{\bf k} (k_z^2 + \gamma_{\bf k}^2)} .
\ee
\end{widetext}
This expression diverges when $\gamma_{\bf k} \to 0$. It happens 
because only the unstable modes are taken into account. The formula 
given in Eq.~(\ref{p2-2-stream-1}) remains finite in the limit 
$\gamma_{\bf k} \to 0$.
The momentum broadening (\ref{qhat-unstable}) grows exponentially 
in time as the spontaneously growing fields exert an exponentially 
growing influence on the propagating parton. This effect is missing 
in the previous results~\cite{Romatschke:2006bb,Baier:2008js} 
obtained for the unstable anisotropic plasma, which was treated as 
a stationary medium. 

The calculation of the momentum broadening  
carried out here does not assume that the medium is 
close to equilibrium or that the growth rate of the instabilities is 
small enough to invoke the static approximation. Even though 
we have restricted our calculation to the simple and analytically tractable 
case of the two stream system, the underlying formalism may be 
applied to calculate the momentum broadening in an arbitrary medium 
as long as the field strength correlators are calculable. 

The differences between the results obtained in this manuscript and 
those in previous articles using a steady state approach are two-fold: 
An unstable system is not time-translation invariant and thus a 
calculation based on time translation invariance is in general invalid.
When the propagator depends independently on $t$ and $t'$, rather 
than on the difference $t-t'$, one obtains a function of two frequency 
variables when performing the (one-sided) Fourier transformations 
with respect to the two time variables.  

In assuming that the gluon occupation number $n(\omega)$ is a function
of only a single frequency, authors of Refs.~\cite{Romatschke:2006bb,Baier:2008js} 
implicitly assume that  the rate of instability growth is slow enough so that 
one may impose approximate time translation invariance over certain time 
periods. Note, however, that even if we ignore the $(t_1+t_2)$-dependence 
in the first term of the correlator (\ref{E^iE^j-2-stream}), the Fourier integral 
with respect to $t_1-t_2$ of the second term does not exist because of the 
exponential divergence. If we were to ignore this problem for the moment, 
we would obtain results which could be reconciled with those of 
Refs.~\cite{Romatschke:2006bb,Baier:2008js} only if we would make the 
additional assumption that the system would remain very close to thermal 
equilibrium. In the following, we illustrate this point in more detail by 
contrasting our calculation with  the results obtained in the 
steady-state approach for the two-stream system. 
It is useful to start from the expression (\ref{qhat-final}) for the 
parameter $\hat{q}$. 

In order to decompose the expression for 
$\hat{q}$ into a convolution of  an occupation number and a 
spectral density we now assume approximate time translation 
invariance. We again restrict our consideration to the 
longitudinal response of the plasma. The calculation is most easily 
carried out in the temporal axial gauge, in which the spectral function 
of longitudinal electric fields is given by
\ba
\label{spec-HTL-TAG}
\rho_L^{ij}(k) = \frac{2\, \Im \varepsilon_L(k)}
{\omega^2 |\varepsilon_L(k)|^2} \: \frac{k^i k^j}{{\bf k}^2} \;.
\ea
A straightforward calculation shows that this expression continues
to hold when the system is anisotropic, but the anisotropy is determined 
by a single vector, here the flow velocity ${\bf u}$. 
The dielectric function $\varepsilon_L(\omega,{\bf k})$ for the two-stream 
system is given in Ref.~\cite{Mrowczynski:2008ae}: 
\begin{widetext}
\ba
\label{eL-3}
\varepsilon_L(\omega,{\bf k}) 
=  1 - \mu^2 
\frac{{\bf k}^2 -({\bf k} \cdot {\bf u})^2}{{\bf k}^2}
\bigg[
  \frac{1}{(\omega - {\bf k} \cdot {\bf u})^2}
+ \frac{1}{(\omega + {\bf k} \cdot {\bf u})^2}
\bigg]
= \frac{\big(\omega^2 - \omega_+({\bf k})^2\big)
\big(\omega^2 - \omega_-({\bf k})^2\big)}
{\big(\omega^2 - ({\bf k} \cdot {\bf u})^2\big)^2} 
\;,
\ea
where $\omega_{\pm}({\bf k})$ is given by (\ref{roots}). We note that 
the expression for $\Im \varepsilon_L(\omega,{\bf k})$ vanishes for real 
$\omega$ and ${\bf k}$ for the two-stream system.  The expression 
$\Im \varepsilon_L(\omega,{\bf k})/|\varepsilon_L(\omega,{\bf k})|^2$
in (\ref{spec-HTL-TAG}) should therefore be replaced by 
\be
- \Im \Big[\frac{1}
{\varepsilon_L(\omega,{\bf k}) + i{\rm sgn}(\omega)0^+} \Big] 
=  \pi \: {\rm sgn}(\omega) \: 
\delta\big(\varepsilon_L(\omega,{\bf k})\big) \;.
\ee
The expression analogous to (\ref{p2-2-stream-1}) derived from the 
steady-state formula (\ref{qhat-final}) then reads:
\ba
\label{qhat-L-1}
\hat q  &=&  - 2\pi g^2 C_R \int \frac{d^4k}{(2\pi)^4} \: n(\omega,{\bf k}) \: 
\delta(\omega-k_z) \, \frac{k_T^2}{{\bf k}^2} \Im \Big[\frac{1}
{\varepsilon_L(k_z,{\bf k}) + i{\rm sgn}(k_z)0^+} \Big] \;.
\nonumber \\
& =& \pi \, g^2 C_R \int \frac{d^3k}{(2\pi)^3} \: n(k_z,{\bf k}) \: 
\frac{k_T^2}{{\bf k}^2} \; {\rm sgn}(k_z) \: 
\delta\big(\varepsilon_L(k_z,{\bf k})\big) \;.
\ea
Here $n(\omega,{\bf k})$ is the average occupation number of the 
soft gluon modes. For the case of a medium in thermal equilibrium, 
$n(\omega,{\bf k})$ is given by the Bose distribution and for soft 
frequencies may be approximated by $T/\omega$. For the case of 
a medium that is barely anisotropic, i.\ e., where the corrections 
to the Bose distribution are small, one may persist with this form 
for the occupation number for short periods of time. 
This is the approximation made in Refs.~\cite{Romatschke:2006bb,Baier:2008js}. 
Such an approximation will, no doubt, break down as the medium continues 
to depart from isotropic equilibrium or as the unstable modes grow with 
time. Besides the insistence on the static approximation, the choice of 
the equilibrium Bose distribution without specifying the short range of 
time over which this is applicable, constitutes an additional source of 
error in these earlier works. 

\end{widetext}

In order to obtain the correct form of the occupation number for any 
time, in the limit that one may still invoke the static approximation, 
one needs to recalculate the two unordered space-time propagators, 
\begin{eqnarray}
D^> (t) &=& \sum_n  \langle n | \hat{\rho}  A (t)  A(0) | n \rangle , 
\nn \\
D^< (t) &=& \sum_n  \langle n | \hat{\rho}  A (0)  A(t) | n \rangle,
\end{eqnarray}
for the specific density operator characteristic to the unstable system 
being considered. In the equation above, $A(t)$ is the vector potential 
of the glue field suppressing Lorentz and color indices, as well as space
variables. In this case one may obtain the spectral density 
(\ref{spectral}) in momentum space and thus calculate the time 
dependent average occupation number as
\begin{eqnarray}
n(\omega) = \frac{D^< (\omega) }{ D^> (\omega) - D^< (\omega) }.
\end{eqnarray}
Only in the case that the density operator is given by the thermal 
Gibbs expression $\hat{\rho} = \exp[ - \beta \hat{H} ] $, will there 
be a simple Kubo-Martin-Schwinger (KMS) relation (\ref{KMS}) resulting 
in the Bose distribution for the occupation number. For any other form 
of the density operator, the occupation number may be considerably 
different from the Bose distribution, even under the assumption of 
approximate time translation invariance. 

As a result, we find that the use of a static approximation and 
a Bose distribution for the occupation number of the unstable modes 
restricts the applicability of the calculations of 
Ref.~\cite{Romatschke:2006bb,Baier:2008js} to very short times 
in a plasma where the growth rates of the instabilities are rather 
small. It should come as no surprise that in such a regime, the effect 
of such modes on the momentum broadening of a hard parton is minimal. 
In an arbitrary weakly coupled quark-gluon plasma with a large momentum 
anisotropy, the growth rate of the unstable modes may be large and, as 
a result, the calculations in the current article remain the only 
consistent approach to calculating the broadening of a hard parton 
traversing such a system. Furthermore, we note that the badly singular 
contributions encountered by the authors of 
Ref.~\cite{Romatschke:2006bb,Baier:2008js} in their calculations
presumably arise from the improper formulation of the problem as 
explained above.

\section{Summary and Outlook}

We have studied the effect of color field fluctuations in a QCD 
medium on the propagation of a hard parton. In the first part of 
this work, we described a classical Langevin approach to compute 
the transport coefficient $\hat{q}$ which denotes the amount of 
transverse momentum broadening per unit path length of an energetic 
parton. In this Langevin approach the coefficient is expressed through 
the correlation function of chromodynamic fields along the light-cone or, 
equivalently, through the unordered gluon propagator. In the case of a 
perturbative equilibrium plasma when the field correlation functions
or the gluon propagators are described in the HTL approximation
we reproduced the known result \cite{Arnold:2008vd,Baier:2008js}. 

In the second part of this study we considered a hard parton in an 
off-equilibrium, unstable plasma. Our explicit calculations performed
for the two-stream plasma showed that unstable modes with exponentially 
growing amplitude cause an exponential growth of the transverse 
momentum broadening. Our result demonstrates that the quasi-static 
approximation used in previous calculations of the transport 
coefficient $\hat{q}$ is not applicable. 

It should be clearly stated, however, that the temporal evolution 
of anisotropic plasma is not the only feature which needs
to be taken into account when transport properties of the 
unstable plasma are studied. When the plasma is populated with 
strong fields, new interaction mechanisms appear. For example, 
it has been recently argued that in a spontaneously chromomagnetized 
plasma the viscosity acquires an anomalous contribution 
\cite{Asakawa:2006tc} while synchrotron radiation contributes to the 
parton energy loss \cite{Shuryak:2002ai,Zakharov:2008uk}.

The formalism for the calculation of $\hat{q}$ used in this work 
is applicable to numerical simulations. It would be interesting 
to perform such a calculation for the case of a realistic, 
three-dimensional quark-gluon plasma off equilibrium. We note that
the calculations of the field correlators are very complex even
in the linear response regime \cite{Mrowczynski:2008ae}.
When the growth of a continuum of unstable modes results in 
strong nonlinear interactions among them, as demonstrated in the 
case of the anisotropic plasma by numerical simulations 
\cite{Arnold:2005ef,Dumitru:2006pz}, an analytical treatment 
of the field correlators seems to be impossible. In view of the 
great phenomenological importance of the transport coefficient 
$\hat{q}$ in relativistic heavy ion physics, a numerical 
evaluation of the field correlator in an unstable, anisotropic 
plasma would be very desirable. According to the exploratory 
simulations \cite{Dumitru:2007rp,Schenke:2008gg}, $\hat{q}$ can 
be significantly larger in the unstable plasma than in an
equilibrium plasma.

Another promising direction for future study is the extension of our
formalism to momentum broadening in a strongly coupled plasma.
Due to the asymptotic freedom of quantum chromodynamics, a 
highly virtual, energetic parton can be weakly coupled to the 
quark-gluon plasma, even if the dynamics of the plasma itself is 
governed by strong coupling. Such an investigation would, for
example, pursue the question of potentially large logarithmic corrections 
to $\hat{q}$ arising from large ratios of scales, such as $Q^2/T^2$, 
where $Q^2$ denotes the virtuality of the hard parton. Instead of relying
on an expansion in orders of the coupling constant $g$, an approach
based on factorization into nonperturbative matrix elements of the medium
and perturbative dynamics of the parton itself  would lend itself 
to a renormalization group treatment \cite{Majumder:2009zu}. Although
$\hat{q}$ would not be calculable {\em ab initio} in such an approach,
its scale evolution with the energy or virtuality of the fast parton 
could be predicted.

\section*{Acknowledgments}

We thank Rolf Baier for helpful correspondence. This work was 
supported in part by the U.S. Department of Energy under grant numbers  
DE-FG02-05ER41367 and DE-FG02-01ER41190. St.M. thanks the members of 
the QCD theory group at Duke University for their warm hospitality 
during his visit.


\end{document}